\documentclass[]{aa}  
\usepackage{graphicx}
\usepackage{txfonts}
\usepackage{natbib}
\usepackage{ulem}
      \def\new#1 {{\bf #1 }}
      \def\cut#1 {\sout{#1} }

\newcommand{\um}{\mbox{\,$\mu$m}}
\newcommand{\oi}{[O{\sc i}]}
\newcommand{\oiii}{[O{\sc iii}]}
\bibpunct{(}{)}{;}{a}{}{,} 

\begin{document}
   \title{A multi-wavelength study of a double intermediate-mass protostar -- from large-scale structure to collimated jets}

\author{J. Forbrich\inst{1,2,3} \and Th. Stanke\inst{4,5}  \and R. Klein\inst{6,7} \and Th.~Henning \inst{8} \and K.~M. Menten\inst{1} \and K. Schreyer\inst{2} \and B. Posselt\inst{6,2,3}}

\offprints{J. Forbrich\\
\email{jforbrich@cfa.harvard.edu}}

\institute{Max-Planck-Institut f\"ur Radioastronomie, Auf dem H\"ugel 69, D-53121 Bonn, Germany 
\and Astrophysikalisches Institut und Universit\"ats-Sternwarte Jena, Schillerg\"a{\ss}chen 2-3, D-07745 Jena, Germany
\and Harvard-Smithsonian Center for Astrophysics, 60 Garden Street, MS 72, Cambridge, MA 02138, U.S.A.
\and Institute for Astronomy, University of Hawaii, 2680 Woodlawn Drive, Honolulu, HI 96822, U.S.A.
\and European Southern Observatory, Karl-Schwarzschild-Str. 2, D-85748 Garching bei M\"unchen, Germany
\and Max-Planck-Institut f\"ur extraterrestrische Physik, Giessenbachstr. 1, D-85748 Garching bei M\"unchen, Germany
\and Department of Physics, University of California, Berkeley, CA 94720, U.S.A.
\and Max-Planck-Institut f\"ur Astronomie, K\"onigstuhl 17, D-69117 Heidelberg, Germany }

   \date{Received / Accepted}

   \abstract{}{The earliest stages of intermediate- and high-mass star formation remain poorly understood. To gain deeper insights, we study a previously discovered protostellar source that is deeply embedded and drives an energetic molecular outflow.}{The source, UYSO\,1, located close to IRAS\,07029--1215 at a distance of about 1\,kpc, was observed in the (sub)millimeter and centimeter wavelength ranges, as well as at near-, mid-, and far-infrared wavelengths.}{The multi-wavelength observations resulted in the detection of a double intermediate-mass protostar at the location of UYSO\,1. In addition to the associated molecular outflow, with a projected size of 0.25~pc, two intersecting near-infrared jets with projected sizes of 0.4~pc and 0.2~pc were found. However, no infrared counterparts to the driving sources could be detected in sensitive near- to far-infrared observations (including \textsl{Spitzer}). In interferometric millimeter observations, UYSO\,1 was resolved into two continuum sources with high column densities ($>10^{24}$~cm$^{-2}$) and gas masses of 3.5~$M_\odot$ and 1.2~$M_\odot$, with a linear separation of 4200~AU. We report the discovery of a H$_2$O maser towards one of the two sources. Within an appropriate multi-wavelength coverage, the total luminosity is roughly estimated to be $\approx$50~$L_\odot$, shared by the two components, one of which is driving the molecular outflow that has a dynamical timescale of less than a few thousand years. The jets of the two individual components are not aligned.  Submillimeter observations show that the region lacks the typical hot-core chemistry.}{We find two protostellar objects, whose associated circumstellar and parent core masses are high enough to suggest that they may evolve into intermediate-mass stars. This is corroborated by their association with a very massive and energetic CO outflow, suggesting high protostellar accretion rates. The short dynamical timescale of the outflow, the pristine chemical composition of the cloud core and absence of hot core tracers, the absence of detectable radio continuum emission, and the very low protostellar luminosity argue for an extremely early evolutionary stage.}


    \authorrunning{Forbrich et al.}
    \titlerunning{A multi-wavelength study of a double intermediate-mass protostar}
   \maketitle
%

\section{Introduction}

In the investigation of the earliest stages of intermediate- and high-mass star formation, the relative
importance of formation processes as different as disk accretion and
coalescence is still an open question (for a recent review, see
e.g.~\citealp{beu07}). Detailed multi-wavelength studies are needed to illuminate the often confusing observational picture. 

Since high-mass protostars are less ubiquitous than their low-mass counterparts, they are on average more distant. The resulting low linear resolution often leads to an oversimplified picture. This is well illustrated by the study of \citet{bsg02}, which reported a case where a single bipolar molecular outflow discovered previously with single-dish radio telescopes in a massive-star--forming (MSF) region was resolved into a multiple outflow system in interferometric data (see also \citealp{bss03}). \citet{dav04} analyzed near-infrared data of two collimated jets in a similar MSF region. They conclude that these jets are very similar to their low-mass counterparts.

We report the results of a comprehensive multi-wavelength follow-up study of a candidate massive protostellar source that was discovered close to IRAS~07029--1215 during a millimeter continuum survey of the surroundings of luminous IRAS sources in the outer galaxy \citep{kle05}. UYSO\,1 is a deeply embedded, very young source at a distance of only 1~kpc, powering a high-velocity bipolar CO outflow; \citet{jan04}\defcitealias{jan04}{Paper~I}\citepalias{jan04} derived the mass of the UYSO\,1 cloud core from the CO(3-2) line emission to be 44~$M_{\odot}$, inferring the hydrogen column density from the integrated CO emission, while the mass derived from optically thin dust continuum emission ($\lambda=850\um$), tracing high column densities, is 15~$M_{\odot}$. The mass of the outflow was estimated to be 5.4~$M_{\odot}$. For details on these earlier mass estimates as well as for a discussion of the distance, we refer to \citetalias{jan04}. Compared to other massive molecular outflows, this mass is at the lower end of the range reported by \citet{bss02}, but most of their sources are much more distant, at an average distance of $4.7\pm3.4$~kpc.
No plausible driving source could be identified in IRAS and MSX mid- and far-infrared data. In single-dish observations, \citet{beu08} detected C$_2$H submillimeter emission towards UYSO\,1, a transition that may be a tracer of the earliest stages of (massive) star formation. 

The aim of our follow-up observations of UYSO\,1 was to study the region at higher angular resolution and to search for the driving source of the enormous CO outflow. For these purposes, we conducted observations in the infrared regime as well as in the millimetric to centimetric wavelength ranges. We note that for very early evolutionary stages, outflow mass entrainment rates and column densities are more conclusive in determining whether a massive star forms than a mass determination since material is still rapidly accreted.
In Section~\ref{sec_obse}, we describe the variety of observations carried out before presenting the results in Section~\ref{sec_resu}. We discuss our findings in Section~\ref{sec_disc} and conclude in Section~\ref{sec_conc}.

\begin{figure}
\centering
\includegraphics[width=\linewidth, angle=-90]{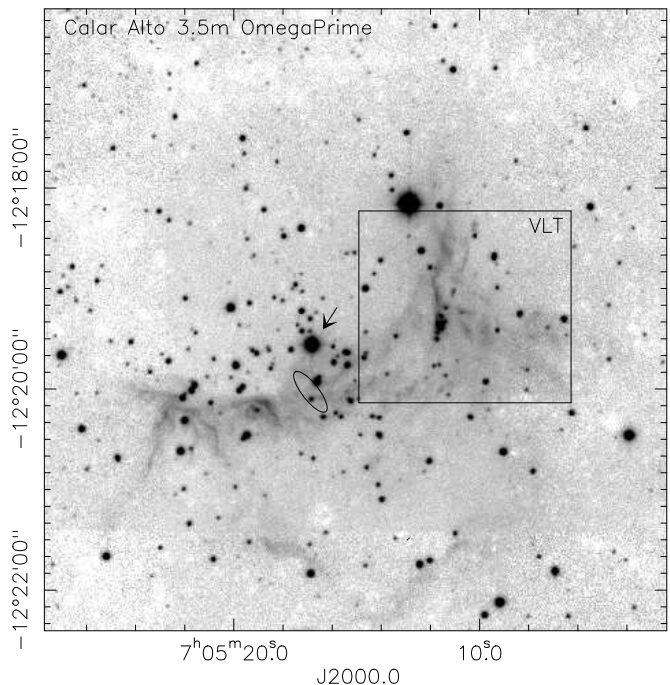}
\caption{H$_2$ S(1) emission, as observed with OmegaPrime at the Calar Alto 3.5m telescope (filter NB2122), in logarithmic scaling. The box denotes the close-up VLT view in Fig.~\ref{jcmtisaac}. The B star HD\,53623 is marked by an arrow and the location of IRAS 07029-1215 is marked by its uncertainty ellipse.}
\label{calar_vlt}
\end{figure}

\begin{figure*}
\centering
\includegraphics[width=17cm, angle=-90]{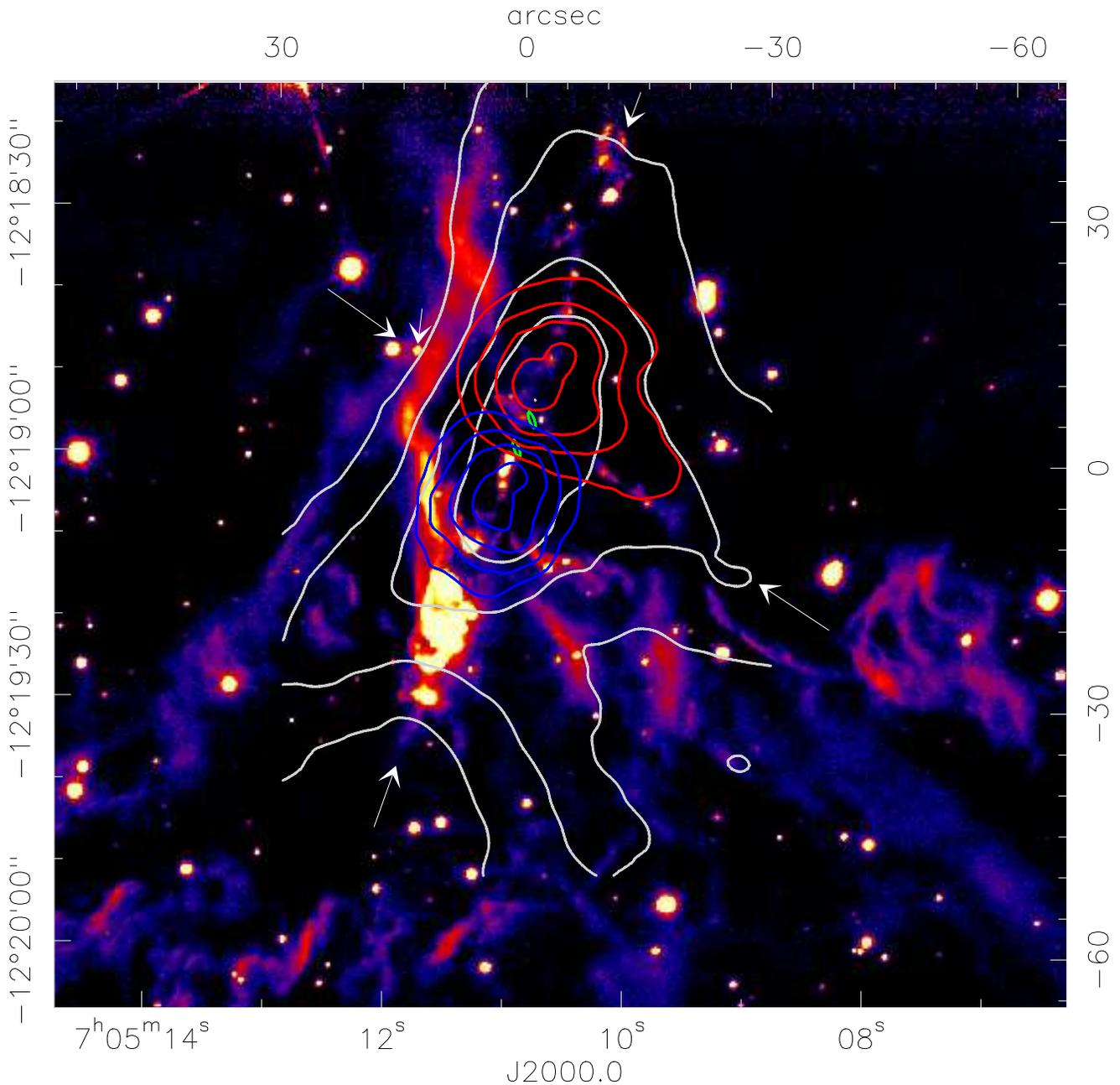}
   \caption{H$_2$ S(1) emission (NB2.13) in logarithmic scaling, overlaid with the JCMT CO(3-2) observations -- total line 10\% to 90\% in 10\% steps, line wings 30\%, 50\%, 70\% and 90\% -- as well as the two continuum sources seen with the PdBI (green ellipses indicating the FWHM synthesized beam). Four arrows indicate the locations of the jets, and an additional small arrow indicates the eastern bow shock of the smaller jet.}
      \label{jcmtisaac}
\end{figure*}

\begin{figure*}
\centering

\includegraphics*[width=\linewidth]{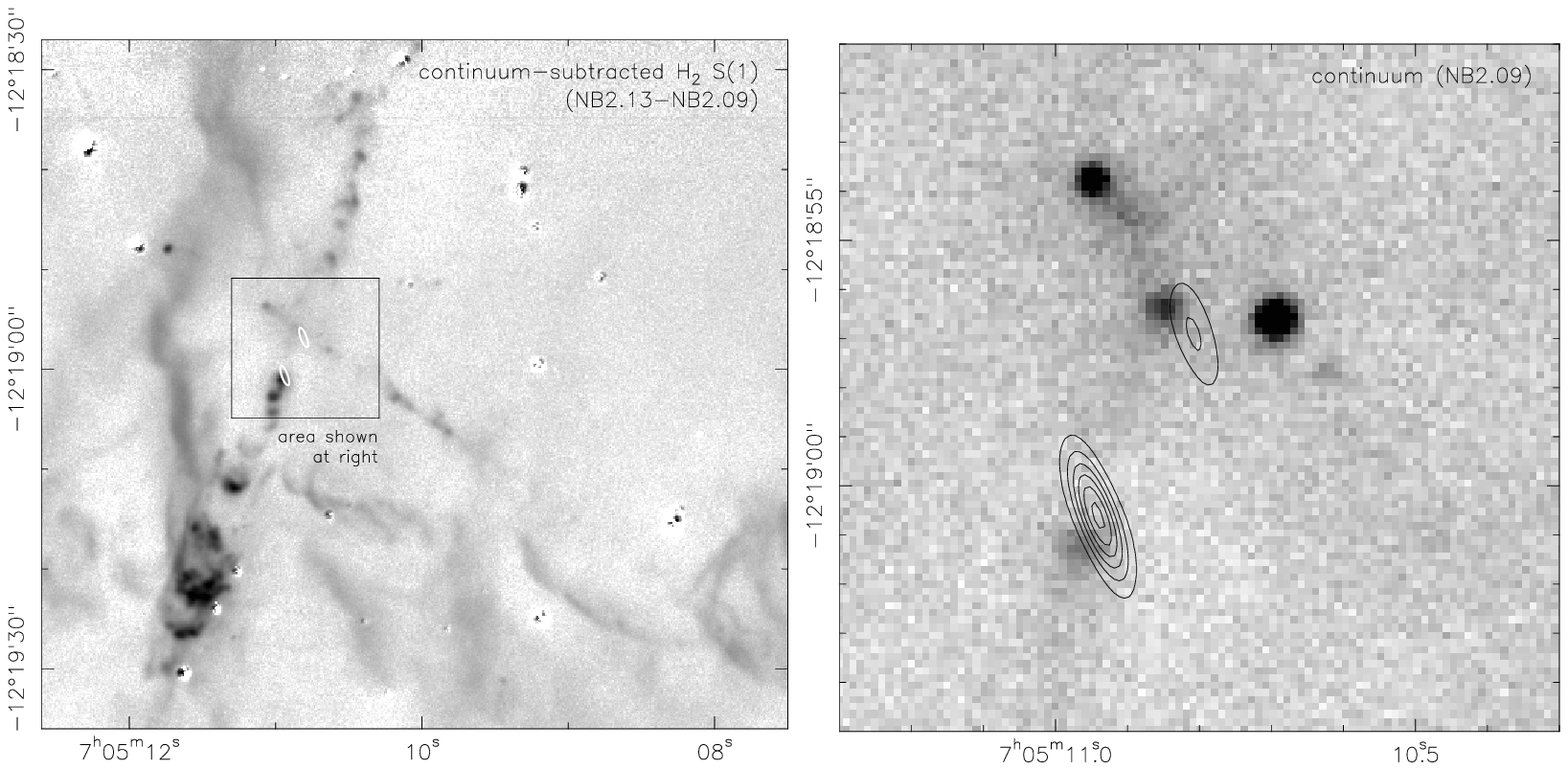}

   \caption{Left: continuum-subtracted H$_2$ S(1) emission (NB2.13--NB2.09) in logarithmic scaling. The subtraction of the brightest stars left a few artefacts. The box indicates the field of view of the right panel; ellipses indicate the positions of the two millimeter continuum sources as in Fig.~\ref{jcmtisaac}. Right: NIR-continuum--only (NB2.09) close-up view of the central area in linear scaling with the 3\,mm PdBI continuum data shown as contours in steps of 1\,mJy.}
  \label{nircont}
\end{figure*}

\section{Observations and data analysis}
\label{sec_obse}

\paragraph{Near- and mid-infrared observations}

In the near infrared, observations were first performed with the 3.5m telescope on Calar Alto, Spain, using the OmegaPrime camera on March 5, 2004, both in $K'$ broadband and H$_2$ S(1) narrowband (NB2122). After a tentative discovery of \textsl{two} jets, intersecting at the position of UYSO\,1, follow-up observations were conducted on several occasions between January 10 and March 1, 2005 with the Infrared Spectrometer and Array Camera (ISAAC) at the Very Large Telescope (VLT) of the European Southern Observatory (ESO). These observations included $K_s$, NB2.09 and NB2.13 as well as \textsl{L}-band imaging. In January and February 2006, UYSO\,1 was observed with the mid-infrared instrument VISIR at the VLT in the PAH2 filter centered on $\lambda = 11.3\um$, with an estimated $3\sigma$ point-source sensitivity of 9~mJy (using the VISIR Exposure Time Estimator v3.2.1). The Calar Alto data were reduced using IRAF while we used the ESO Eclipse 5.0 software for the VLT data.

\paragraph{Far-infrared observations}

We used the Multiband Imaging Photometer for Spitzer \citep[MIPS,~][]{rie04} onboard the \textsl{Spitzer} Space Telescope to obtain maps at 24{\um} and 70{\um} as well as a low-resolution spectrum from 52{\um} to 105{\um} in the SED mode. The observations were executed on November, 11 and 12, 2005. The imaging data presented here are post-BCD maps processed with version S16.1 of the pipeline. Inspection of the post-BCD data products and manual reduction with the mopex software package showed that the pipeline reduction of the data is reliable in our case. The widespread extended emission made estimating the background difficult. Some striping along the detector columns remain in the 70{\um} maps. For display, the stripes were reduced by smoothing with a three-pixel median filter aligned perpendicularly to them. Measurements, however, were done on the unsmoothed maps.
The MIPS-SED data were reprocessed with mopex, starting with the BCD products created by the pipeline (S16.1.1) to create the spectra along the slit.

\begin{table}[btp]
  \caption{Table of relevant positions$^{\rm a}$ (Epoch 2000)}
  \centering
  \begin{tabular}{lll }
    \hline\hline
    Name&\multicolumn{1}{c}{RA (h:m:s)}&\multicolumn{1}{c}{DEC ($^\circ$:':'')}\\
    \hline
    \hline
    UYSO\,1a          &07:05:10.940(1)&-12:19:00.64(3)\\
    UYSO\,1b          &07:05:10.811(3)&-12:18:56.84(9)\\
    H$_2$O maser      &07 05 10.8105(1)&-12:18:56.807(3)\\
    1st 70{\um} peak  &07:05:10.96&-12:19:09.9\\
    2nd 70{\um} peak  &07:05:11.09&-12:19:33.1\\
    24{\um} peak      &07:05:11.56&-12:19:19.4\\
    IRAS 07029--1215  &07:05:16.9 &-12:20:02 \\
    \hline
    \hline
  \end{tabular}
  
  \flushleft{$^{\rm a}$ Numbers in brackets denote the uncertainty in the last digit.}
  \label{tab:pos}
\end{table}

\paragraph{Millimeter and submillimeter observations}

Since the discovery observation of the CO outflow, two considerably larger and fully sampled CO(3-2) maps of the region were taken on June 4 and 6, 2005, with the facility receiver B3 at the James-Clerk-Maxwell Telescope (JCMT). In December 2003, a map of UYSO\,1 and its surroundings in CO(2-1) was taken with the Heterodyne Receiver Array HERA at the IRAM 30m telescope. However, since the CO(3-2) data cover a larger area at a better signal-to-noise ratio, we focus our analysis on these data. In July 2005, N$_2$H$^+$(1-0), N$_2$D$^+$(1-0), HCO$^+$(3-2) and DCO$^+$(2-1) were observed towards UYSO\,1 with the IRAM 30m telescope and its facility receivers in frequency switching mode.

After first, very short observations with the IRAM Plateau de Bure Interferometer (PdBI) in its low-angular resolution D configuration in September/October 2003, UYSO\,1 was reobserved in the newly extended high-resolution A configuration in February 2006. While the synthesized beam size in the former observations was $13\farcs9\times5\farcs9$ for the 3\,mm continuum, this improved to $2.2'' \times 0.6''$ in the latter ($0.85'' \times 0.23''$ at 1~mm). Besides the continuum, the CS(2-1)  and CO(2-1) transitions were observed. In order to be able to account for short-spacings in the $uv$ plane, the source was previously mapped in both transitions with the IRAM 30m telescope. 

In November 2005, UYSO\,1 was observed with the Caltech Submillimeter
Observatory (CSO) on four occasions (Nov 17/18, Nov 19, Nov 20, Nov 22), looking for several submillimeter transitions in the 345~GHz and 220~GHz ranges. The beam sizes\footnote{http://www.submm.caltech.edu/cso/receivers/beams.html} (FWHM) of the CSO are $\sim22''$ and $\sim31''$ in the 345~GHz and 220~GHz bands, respectively.
In the same month, the source was also observed with the Atacama Pathfinder Experiment (APEX) telescope \citep{gus06} to obtain a better CO(3-2) spectrum at the source position. The FWHM beam size\footnote{http://www.apex-telescope.org/telescope/} of the APEX telescope at 345~GHz is $\sim18''$.
All data from the 30m, APEX, and the PdBI were reduced using the GILDAS software developed by IRAM and Observatoire de Grenoble. 

\paragraph{Centimeter radio observations}

In November 2003, H$_2$O maser emission towards UYSO\,1 was discovered with the Effelsberg 100m radio telescope. The discovery was confirmed with the same telescope in March 2004. At the same time, as well as in January 2004, the source was searched for NH$_3$(1,1) emission with the Effelsberg 100m telescope. On April 13, 2005, it was searched for signs of CH$_3$OH maser emission, using the Toru\'n 32m radio telescope.

In March 2006, high-resolution radio observations were carried out with the NRAO Very Large Array (VLA) in A configuration. Both the 8.4~GHz radio continuum emission as well as the newly discovered H$_2$O maser emission were studied. The Effelsberg data were reduced using the IRAM GILDAS software, and for the VLA data, we used the NRAO AIPS software.

\begin{figure*}
\centering
\includegraphics[width=\linewidth]{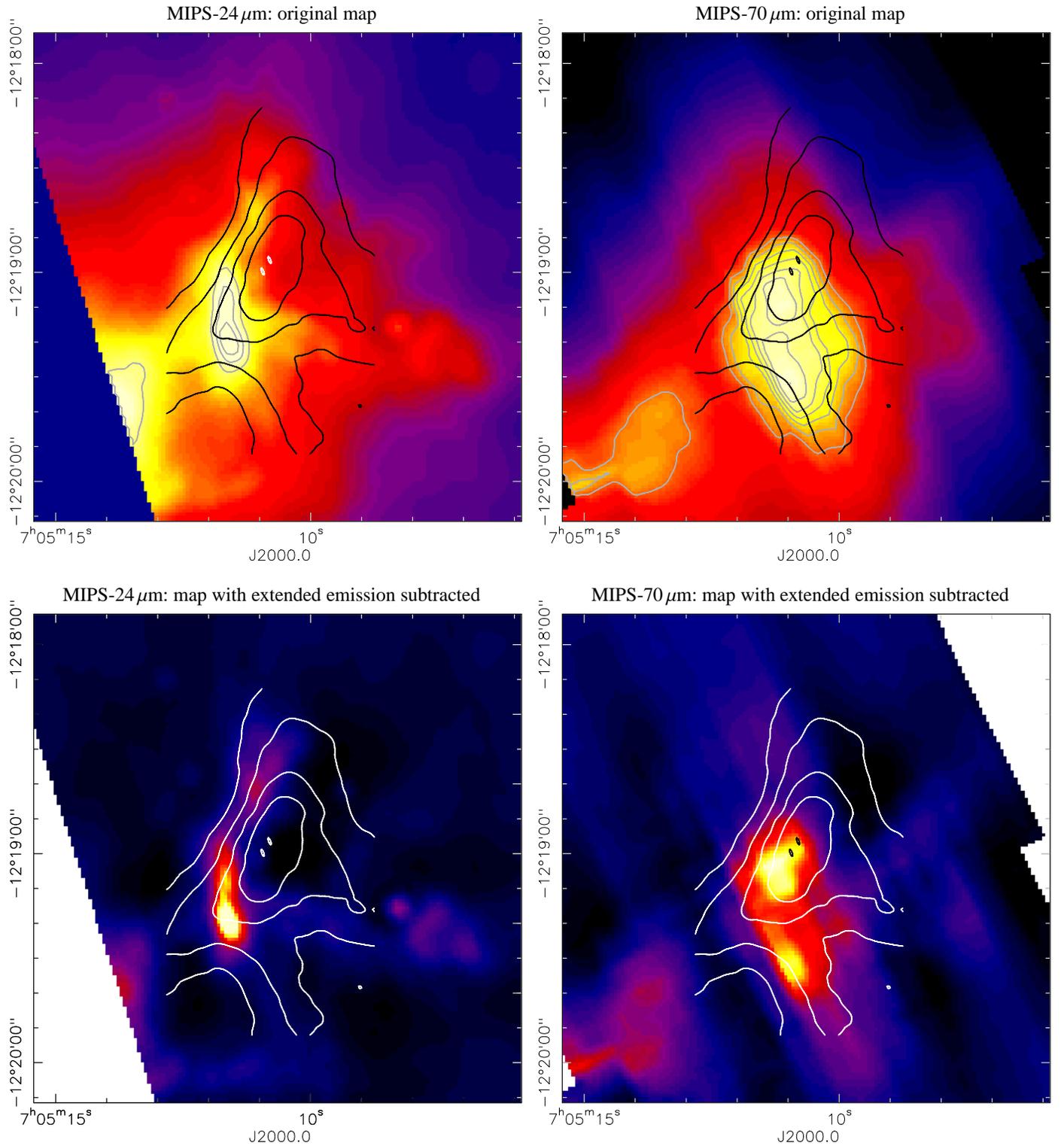}
\caption{MIPS 24{\um} (left) and 70{\um} (right) data. Upper panel: original maps, in logarithmic scaling. Contours of the data (grey) and of the CO(3-2) emission (black) are displayed as reference. Lower panel: maps with extended emission filtered out (see text), in linear scaling. Contours of the CO(3-2) emission are displayed as reference. The color scale is different from the one used in the upper panel. In all plots, the locations of UYSO\,1a and UYSO\,1b are marked with small ellipses, as in Fig.~\ref{jcmtisaac}.}
  \label{uyso1mips}
  \label{uyso1mips2}
\end{figure*}

\section{Results}
\label{sec_resu}

Based on the new multi-wavelength data, we report the discovery of two collimated jets that appear to be connected to two previously unresolved protostars at the location of UYSO\,1. In spite of deep searches, we did not detect any infrared counterparts of these protostars. Towards one of the two protostars, water maser emission was discovered. Observations of molecular transitions reveal that UYSO\,1 does not show typical hot-core chemistry. In the following, we present these results in more detail, differentiating large-scale structure (the clump in which UYSO\,1 is embedded, the outflow, and the jets) and small-scale structure (the protostars UYSO\,1a and UYSO\,1b) before discussing the spectral energy distribution (SED). Relevant positions are listed in Table~\ref{tab:pos}.

\subsection{Large-scale structure}
\label{ssec:largescale}

\paragraph{Near- and mid-infrared observations}
The initial discovery of two collimated jets, obtained with the Calar Alto 3.5m telescope (Fig.~\ref{calar_vlt}), was clearly confirmed by the VLT observations (Fig.~\ref{jcmtisaac}). The larger field of view of the Calar Alto narrowband image additionally shows H$_2$ S(1) emission in the outer parts of the H\,{\sc II} region Sh2-297 which is powered by HD~53623. The two jets intersect with an angle of 75$^\circ$ at about the submillimeter continuum position of UYSO\,1, a first indication that UYSO\,1 may harbor more than one source. The larger north-south jet has an apparent size of 1.3', corresponding to 0.4~pc at a distance of 1~kpc. It also has multiple bow shocks, best seen in the continuum-subtracted image in Fig.~\ref{nircont}. This jet deviates considerably from a straight line in its northern lobe, possibly a sign of interaction with the surrounding medium.
The smaller jet, terminating in a bright, very compact bow shock at its eastern end, is at least 0.2\,pc long (it might be even longer, as a barely visible chain of features extends well beyond the bright section of its western lobe).
Both jets have a knotty structure. A quantitative estimate of the inclination angles of the two jets with respect to the line of sight is difficult. However, the fact that we clearly see both lobes of both jets in spite of the surrounding  material suggests that they lie close to the plane of the sky. The narrowband luminosities of the two jets, as determined from the 2.12~$\mu$m data, are $L_{\rm 2.12\mu m}=0.02L_\odot$ for the larger, north-south jet and $L_{\rm 2.12\mu m}=0.0007L_\odot$ for the smaller, east-west jet. The luminosity of the larger flow is at the upper end of the range of values found in a survey of Orion A \citep{sta02}, a first suggestion that the driving source is at least of intermediate mass.

Another prominent feature is the ridge of strong H$_2$ emission east of the
core, extending from the north to the south, also visible in the Calar Alto
image. This ridge appears to be the edge of the molecular clump where UYSO\,1 is
embedded in since also the CO and millimeter continuum emission drop dramatically in that region. The emission ridge indicates where the H\,{\sc II} region is interfacing the molecular cloud. Whether the H$_2$ emission is excited by shock fronts in the ionisation front or by UV pumping in the photon-dominated region (PDR) between the H\,{\sc II} region and the molecular cloud, we cannot decide, although a PDR appears more likely (see the discussion of oxygen fine structure lines in Section \ref{ssec:photo}). This would require the comparison of the relative intensities of several H$_2$ lines \citep[e.g.,][]{hol77}, data that are not available at present. The H$_2$ emission in jets is usually shock-excited.

\paragraph{Far-infrared observations}

Neither 24\,$\mu$m nor 70\,$\mu$m mid-infrared sources were found to be in
direct relation to the millimetre continuum peak (Fig.~\ref{uyso1mips}), even though the 70\,$\mu$m observations may show some emission at that position (see discussion of the SED in Section~\ref{ssec:photo}). Instead, the 24\,$\mu$m emission closely follows the H$_2$ S(1) emission ridge as the PDR is also heating the dust at the surface of the clump. The 70\,$\mu$m
emission also starts at this rim, but extends further into the cloud as it
traces colder material deeper in the cloud. The emission peak in the
70\,$\mu$m image is separated by 13{\arcsec} to the north-west from
the 24\,$\mu$m peak. It is still another 11{\arcsec} away from the millimeter
sources. In Sect. 3.3, we discuss the possibility of UYSO\,1 being swamped by
surrounding extended emission and try to determine its potential contribution
to the FIR emission.

\begin{figure}
\centering
\includegraphics*[width=5cm, angle=-90, bb= 180 60 600 760]{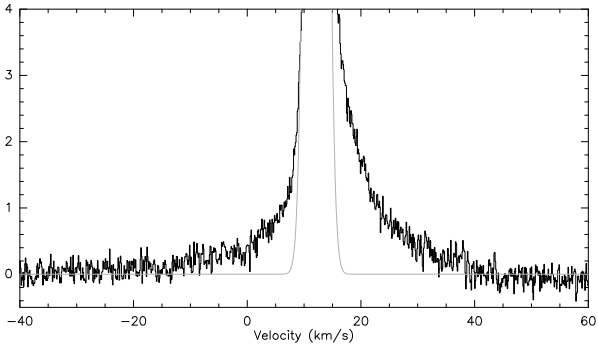}
\caption{CO(3-2) spectrum towards UYSO\,1, as observed with the APEX telescope, in $T_a^*$~[K]. In order to show the line wings, only the lower 10\% part of the line are shown. The grey line shows a Gaussian profile fitted to the entire line.}
  \label{apex_uyso1_sum}
\end{figure}

\begin{figure}
\centering
\hspace*{7.7cm}
\includegraphics*[width=8cm, angle=-90, bb= 54 575 542 20]{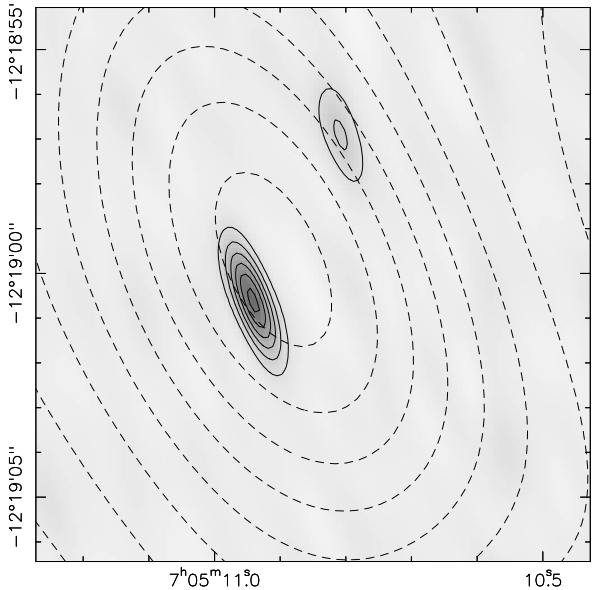}
   \caption{The two 3\,mm continuum observations using the PdBI, at the same angular scale (greyscale and solid contours: A configuration, contour step: 1\,mJy; dashed contours: D configuration, contour step: 2\,mJy). The brighter source is UYSO\,1a, the fainter one is UYSO\,1b.}
  \label{uyso1pdbi1}
\end{figure}

\begin{figure}
   \centering
   \includegraphics[width=4.7cm, angle=-90]{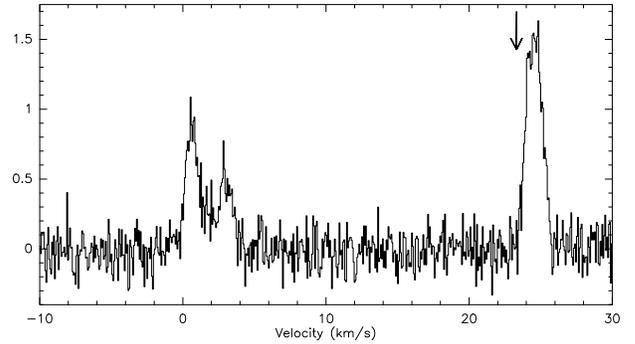}
   \caption{H$_2$O maser emission towards UYSO\,1, as observed with the Effelsberg 100m telescope, axis units are Jy. The VLA detection of an H$_2$O maser in UYSO\,1b has an unresolved velocity of 23.3~km\,s$^{-1}$, marked by an arrow, with a peak flux density of 0.92~Jy. The system velocity is v=12~km\,s$^{-1}$.}
   \label{uyso1eberg}
\end{figure}

\paragraph{(Sub-)millimeter and centimeter radio observations} 
The new CO(3-2) single-dish map, shown in Fig.~\ref{jcmtisaac}, corroborates the results presented in \citetalias{jan04}. There is a strong gradient in emission towards the H\,{\sc \small II} region east of the position of UYSO\,1. The APEX CO(3-2) spectrum taken at the position of UYSO\,1 shows the line wings most clearly (Fig.~\ref{apex_uyso1_sum}). The outflow is prominent in both the CO(3-2) and the CO(2-1) maps. Within the uncertainties, the molecular outflow coincides with the larger of the two NIR jets. 
Thus, in the single-dish CO maps, one massive outflow already discussed in \citetalias{jan04} is detected, but the new, larger CO(3-2) map additionally shows an indication of an east-west outflow, at least at its western end. Possibly, the millimeter outflow related to the smaller jet is weaker than the large previously known outflow and as such difficult to detect when both systems are superimposed.

From the CO(3-2) map, the mass of the outflow can be estimated in direct comparison to \citetalias{jan04}, following the procedure outlined there as well as in \citet{hen00}, using the proportionality of the integrated CO main-beam temperature and H$_2$ column density. The new, larger map has a slightly smaller SNR than the data used in \citetalias{jan04}, affecting the line wings. For the entire line profile, the 40\% intensity contour (which is not entirely within the map) traces 33~$M_\odot$, the red- and blueshifted line wings trace 2~$M_\odot$ and 1~$M_\odot$, respectively (down to the 10\% contour). These numbers are slightly lower than those derived from the deeper data in \citetalias{jan04}, where the cloud mass was estimated to be 40~$M_\odot$. While it is difficult to give quantitative uncertainties for these mass estimates, we note that the masses are uncertain by at least a factor of a few, i.e., they are order-of-magnitude estimates.
While for the extent and the maximum velocities of the outflows, the original JCMT data in the outflow lobes are still the best due to their SNR, we note that our deeper APEX pointing on UYSO\,1 shows line wings that are comparable to what was previously only detectable in the outflow lobes. The outflow velocity $v_{\rm proj}=30$~km\,s$^{-1}$ from \citetalias{jan04} compares to $v_{\rm proj}=26.5$~km\,s$^{-1}$ derived at the position of UYSO\,1 (Fig.~\ref{apex_uyso1_sum}, compare to Fig.~4 in Paper~I). In the new, large CO(3-2) map, the 10\% contours of the line wings indicate a total size of the molecular outflow of 52'', or 0.25~pc. Compared to \citetalias{jan04}, the inferred size of the outflow is basically unchanged.
Assuming that we have an edge-on disk geometry rather than a pole-on view (as is suggested by the clear detection of both near-infrared jet lobes) allows us to better constrain the outflow and its mass entrainment rate.
An inclination angle of 80$^\circ$ with respect to the line of sight yields an outflow dynamical timescale ($t_d(i)=R_{\rm out}(i)/v_{\rm out}(i$) of only a few hundred years and an outflow mass entrainment rate of $\dot{M}=4\times10^{-3}\,M_\odot$\,yr$^{-1}$. Even for an inclination of $i=57.3^\circ$ (see Paper I), the dynamical timescale is still only 3000 years with a mass entrainment rate of $\dot{M}=1\times10^{-3}\,M_\odot$\,yr$^{-1}$.
We can conservatively constrain the dynamical timescale to less than $10^4$~yr except for very small inclination angles ($< 25\degr$) and the mass entrainment rate correspondingly to $>3\times10^{-4}\,M_\odot\rm yr^{-1}$: as discussed above, also the NIR observations of the jet lobes suggest large inclination angles with respect to the line of sight. Also, in a pole-on configuration with small inclination angles, we should see the driving sources in the NIR/MIR when looking into the outflow cavities, which is not the case.
Based on these outflow properties, we estimate and discuss the accretion luminosity in Section~\ref{sec_disc}.

New molecular line observations beyond those in \citetalias{jan04} are summarized in Table~\ref{linetab}. To better trace the overall mass, we also studied UYSO\,1 in $^{13}$CO(2-1). Following \citet{sco86}, the emission peak at the position of UYSO\,1 corresponds to a mass of 8~$M_\odot$ in the CSO beam size of $\sim33$'' (FWHM). Interestingly, neither NH$_3$ nor N$_2$H$^+$(1-0) emission was found towards the source, with upper limits of $<$0.15~K and $<0.1$~K, respectively. HCO$^+$(3-2) and DCO$^+$(2-1) were detected. The HCO$^+$(3-2) line may consist of several velocity components. Several submillimeter molecular lines that we searched for using the CSO were not detected (see footnote in Table~\ref{linetab}). These results indicate that UYSO\,1, with only very few detectable molecular transition lines, does not show typical hot-core chemistry. 

Based on the upper limit of N$_2$H$^+$(1-0) emission, we can estimate corresponding upper limits for the column density and the abundance. For the column density, $N_{N_2H^+}\approx 8\times10^{11}\Delta v T_R$~cm$^{-2}$ \citep{ben98}, we derive an upper limit of $N_{N_2H^+}\approx1.6\times10^{11}$~cm$^{-2}$, assuming a line width of 1~km\,s$^{-1}$. An upper limit for the abundance follows when relating this to the hydrogen column density derived by \citet{kle05} from the submillimeter radio data at 850~$\mu$m, $N_H = 5.8\times10^{22}$~cm$^{-2}$ (for $T=20$~K). The result, an abundance of only $N_{N_2H^+}/N_H\le2.6\times10^{-12}$, is surprisingly low compared to values of $\approx 10^{-10}$ found in starless cores by \citet{taf04b}. 

\begin{table}
\caption{Molecular transition lines observed towards UYSO\,1}
\label{linetab}
\centering
\begin{tabular}{lll}
\hline\hline 
Transition & Telescope & $T_a^*$~[K]\\
\hline
NH$_3$(1,1)     & Effelsberg 100m & $<0.15$~K \\ 
N$_2$H$^+$(1-0) & IRAM 30m        & $<0.1$~K \\
HCO$^+$(3-2)    & IRAM 30m        & 3.5~K \\
DCO$^+$(2-1)    & IRAM 30m        & $0.5$~K \\
$^{13}$CO       & CSO$^{a}$       & 1.8~K\\
HCN(4-3)        & CSO             & 0.7~K\\
CCH$^{b}$       & CSO             & $<0.4$~K \\
H$_2$CO(5$_{(1,5)}$-4$_{(1,4)}$) & CSO & 0.3~K \\
\hline
\end{tabular}

\flushleft
$^{a}$ notable \textsl{undetected} molecules in our 345~GHz-range CSO observations: CH$_3$CN, CH$_3$OH, SO, SO$_2$, HNCO, HCOOCH$_3$, HCCCN, CH$_3$CH$_2$CN, and additionally in the 220~GHz range: SiO\\
$^{b}$ blends at 349.338~GHz and 349.400~GHz
\end{table}

\subsection{Small-scale structure}
\label{ssec:smallscale}

\paragraph{Millimeter observations} 
The millimeter continuum data from the two PdBI observations are ideal to study the small-scale dust continuum emission. In the low-resolution (D configuration) data, only a single source is detected at 3\,mm. In the high-resolution data (extended A configuration), this source is resolved into two protostars, UYSO\,1a and UYSO\,1b (Table~\ref{tab:pos}); both sources remain unresolved. At 1\,mm, only the brighter southern source, UYSO\,1a, is clearly detected due to limited sensitivity. Results of the two 3\,mm continuum datasets are shown in Fig.~\ref{uyso1pdbi1}.

We follow \citet{bss03,beu05err} in determining the mass from the 3\,mm continuum data, as traced by the optically thin dust emission of the protostellar envelopes. We assume a temperature of $T= 45$~K \citepalias{jan04} while leaving the gas-to-dust ratio (100:1) and other quantities the same as in \citet{bss03} who estimate the results to be accurate within a factor of five. The single source detected in the low-resolution data corresponds to a gas mass of 9.5~$M_\odot$. In the high-resolution data, two continuum sources are detected in the 3\,mm band. These are at roughly the positions of two faint near-infrared continuum sources, but do not fit exactly (see discussion of the NIR observations below). Both objects appear to be intermediate-mass protostars with gas masses of 3.5~$M_\odot$ for the more massive component, UYSO\,1a, and 1.2~$M_\odot$ for the second component, UYSO\,1b. The linear separation between the two sources is $4\farcs17$, or 4200~AU. The hydrogen column densities corresponding to the derived gas masses are $N_{\rm H}=6.2\times10^{24}$~cm$^{-2}$ and $N_{\rm H}=2.2\times10^{24}$~cm$^{-2}$ for UYSO\,1a and UYSO\,1b, respectively. As a crude approximation, we convert these to visual extinction according to the empirical relation $N_{\rm H}$[cm$^{-2}]\approx 2\times10^{21}\times A_{\rm V}$[mag] \citep{ryt96,vuo03}, resulting in visual extinctions of $A_V>$1000~mag. We note that the derived column densities are well above typical values for low-mass protostars (e.g., \citealp{mot01}) and also above the minimum column density of $3\times10^{23}$~cm$^{-2}$ (or 1\,g\,cm$^{-2}$) that \citet{kru08} derived for the formation of massive stars.

The molecular line data collected with the PdBI suffers from missing flux, and the two extreme configurations are difficult to combine due to virtually no overlap in the \textsl{uv} plane. Thus, we only briefly discuss the CS(2-1) data here. While in the D configuration, again only a single source is detected with barely noticable velocity structure, the A configuration data show emission that is largely resolved out.

\paragraph{Centimeter radio observations} 

H$_2$O maser emission, a signpost of low- and high-mass star formation \citep{hen92}, was discovered in single-dish observations towards UYSO\,1 in two velocity components symmetrically spaced around the system velocity of v$_{\rm lsr}$=12~km\,s$^{-1}$ (Fig.~\ref{uyso1eberg}). One of the two components has a double-peaked substructure. In high-resolution VLA A-array observations, however, only a single maser spot was found at the position of the north-western millimeter continuum source, UYSO\,1b, close in velocity to the brighter component in the Effelsberg maser spectrum. No CH$_3$OH maser emission was detected ($<0.3$~Jy). In the centimeter continuum at wavelengths of 1.3~cm and 3.5~cm, only 3$\sigma$ upper limits of 5.6~mJy and 0.21~mJy for the flux densities could be determined, respectively. Thus, no detectable H\,{\sc II} region has formed yet.

\paragraph{Near- and mid-infrared observations}
In deep NIR imaging, two sources were detected in the NB2.09 continuum and the \textsl{L} band which are close to the positions of the millimeter protostars, but they do not coincide with them (Figs.~\ref{jcmtisaac} and \ref{nircont}). The astrometric accuracy of the near-infrared data, refined with positions from the 2MASS catalogue, is estimated to be $<0.2''$. Given the enormous visual extinction towards the two protostars (see above), we probably only see scattered light from their vicinities \citep[e.g.,][]{lin05,wei06}. Since UYSO\,1a lies in the large north-south jet, and UYSO\,1b lies in the east-west jet, the two millimeter sources probably are their driving sources. Then, UYSO\,1a would also power the dominant molecular outflow coinciding with its jet while UYSO\,1b excites the water maser. Notably, the NB2.09 NIR continuum source just to the south-east of UYSO\,1a and the source just to the east of UYSO\,1b appear not to be point sources; it is also noteworthy that both appear on the sides of the blueshifted outflow lobes, which would be tilted towards us and thus suffer less extinction. In the VISIR observations at 11{\um}, no source was detected at the position of UYSO\,1. As an upper limit, we use the expected $3\sigma$ point source sensitivity of 9~mJy.


\subsection{Spectral Energy Distribution}
\label{ssec:photo}
Compared to the spectral energy distribution of UYSO\,1 in \citetalias{jan04} which was based on the submillimeter detections and upper limits derived from IRAS data, we now have much more comprehensive information to constrain the combined luminosity of UYSO\,1a and 1b. 

\paragraph{Broad-band Spitzer observations}

As noted in Sect.~\ref{ssec:largescale}, UYSO\,1 remains undetected in the MIPS scan maps. Bright
extended emission from the clump around UYSO\,1 is detected but no compact
emission peak that could be ascribed to the core containing UYSO\,1a and
1b is seen. The question is how to estimate upper limits for the flux density of UYSO\,1
at 24\,$\mu$m and 70\,$\mu$m. As a conservative estimate, we perform aperture
photometry at the position of UYSO\,1, using apertures of {14\arcsec} and
{32\arcsec} in diameter with background annuli from {80\arcsec} to
{100\arcsec} and {78\arcsec} to {130\arcsec}, for 24 and
70\,$\mu$m, respectively. These apertures include UYSO\,1 but do not cut into the PDR,
whereas the background annuli lie beyond the main extended emission. We applied the appropriate aperture and color corrections as described in the MIPS Data
Handbook v3.3 as derived from the theoretical PSF.  This method yields
conservative upper limits of 0.3\,Jy and 70\,Jy for UYSO\,1 at 24\,$\mu$m and
70\,$\mu$m, respectively.

However, these values mainly measure the large-scale emission of the externally heated envelope at the position of UYSO\,1. Inspired by unsharp masking, we estimate the extended emission by convolving the maps with a Gaussian profile and then subtracting it to retain the small-scale structure. The FWHM of the
Gaussian has been 40\arcsec{}. The unsharp-masked maps are displayed in the
lower panels of Fig.~\ref{uyso1mips}. The small-scale structure is now much more
pronounced, but no peak at the position of UYSO\,1 becomes apparent. Repeating
the aperture-photometry should now result in values much less affected by the
extended emission. Virtually no flux is left in the 24\,$\mu$m map at the position of UYSO\,1. Only an upper limit of 0.1\,Jy can be derived. At 70\,$\mu$m, the aperture photometry retains $25\pm2$\,Jy at the position of UYSO\,1. This number is a second upper limit since there is no compact emission apparent at this position (although there is a conspicuous extension of the 70\,$\mu$m emission towards the positions of UYSO\,1a and b, possibly indicating that these protostars start to become visible at about this wavelength). There is considerable uncertainty as to how well the extended emission and the cloud background are subtracted.

\begin{figure}[tbp]
  \centering
  \includegraphics[width=\linewidth]{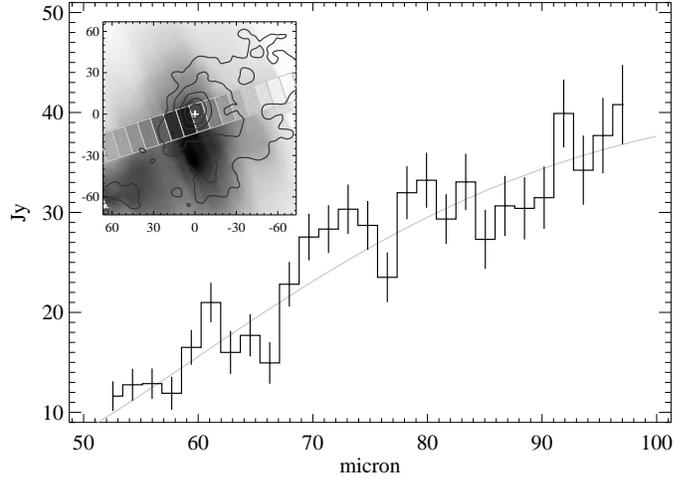}
  \caption{MIPS Spectrum of UYSO\,1 with $3\sigma$ error bars; in grey a 40\,K
    black body spectrum; inset: Location, size, and integrated intensity of
    the MIPS slit on 70\,$\mu$m-map with 850\um-contours}
  \label{fig:sed}
\end{figure}

\paragraph{Low resolution Spitzer spectroscopy (SED mode)}

Even though UYSO\,1 remains undetected in the MIPS imaging data, the low-resolution spectroscopy can help characterizing the surroundings. The inset of Fig.~\ref{fig:sed} shows the location, the size, and the pixels of the MIPS slit relative to the 70{\um} imaging data; the resulting spectrum is shown in the main part of the figure. This spectrum was extracted from the three pixels closest to the position of UYSO\,1. The cloud background was removed by subtracting the average of the three pixels east and three pixels west of the UYSO\,1 spectrum. All the spectra are derived by averaging over three pixels along the slit and applying the proper aperture correction.  The $3\sigma$ error bars are the uncertainties propagated from those reported in the post-BCD products. The spectrum derived for the position of UYSO\,1 is compatible with the flux derived from the 70{\um} map. It does not show any strong features and the slope can be described by a black body of temperature 40\,K. Note again that UYSO\,1 itself is probably not detected.

\begin{figure}[tbp]
  \centering
  \includegraphics[width=\linewidth, bb=37 10 335 310]{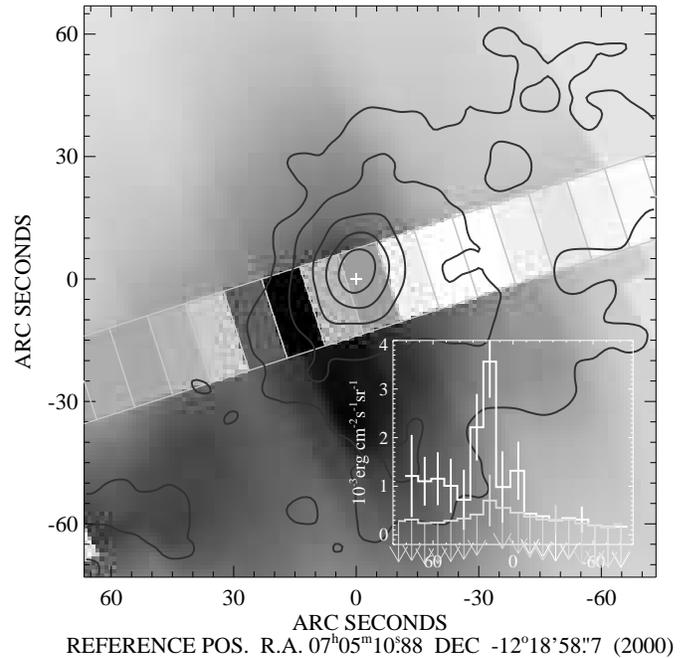}
  \caption{The {\oi} and {\oiii} line: The intensity in the MIPS slit are the
    {\oi} 63{\um}-line on top of the 70{\um}-map (850\um{} contours). The
    inset displays again the variation of the {\oi} 63{\um}-line across the
    slit plus the {\oiii} 88{\um}-line (grey). Error bars are $3\sigma$.}
  \label{fig:oi}
\end{figure}

\paragraph{Oxygen fine structure lines}

A feature that is detected in the \textsl{Spitzer} SED-mode observations despite its low spectral resolution are oxygen fine structure lines at the eastern rim of the molecular core.  Fig.~\ref{fig:oi} again shows the 70{\um} map with the MIPS slit, this time with the intensities of the {\oi} line at 63{\um}. The inset shows the variation of the {\oi} line and the {\oiii} (88{\um}) line along the slit. The subtracted continuum was obtained by averaging the two pixels neighboring the line-containing pixel in wavelength.

We clearly see here a PDR radiating in two important cooling lines on the eastern side of the core where the radiation from IRAS~07029-1215 impacts the molecular cloud. The emission comes only from a small area as atomic oxygen is only present up to optical depths of $A_V\approx10$\,mag, and the PDR must be seen edge on. The {\oi} line's intensity of a few $10^{-3}\rm\,erg\,cm^{-2}s^{-1}sr^{-1}$ is expected for PDRs with densities of $10^3$ to $10^4\rm\,cm^{-3}$ \citep{hol99}.

\section{Discussion}
\label{sec_disc}

In view of our previous knowledge on UYSO\,1, the main new result that warrants discussion is its newly constrained SED. Also, the nondetections in N$_2$H$^{+}$ and NH$_3$ are of special interest.

The non-detection of UYSO\,1 at mid-infrared wavelengths is surprising given the earlier estimate of the SED in \citetalias{jan04}. The mid-infrared part of that SED was only weakly constrained by upper limits from IRAS data, clearly contaminated by the nearby bright source IRAS~07029-1215. 
The previous estimate includes the entire cloud core, heated also externally. The previous luminosity upper limit of $<1900\,L_\odot$ corresponds to the large-scale emission in the \textsl{Spitzer} maps. Between 24 and 70{\um}, the luminosity of the filtered-out large-scale emission is $\sim1500\,L_\odot$.

We can estimate the amount of energy that is intercepted and re-radiated by the PDR neighboring UYSO\,1. The luminosity estimate in \citetalias{jan04} includes any such emission inside the diameter of 80\% encircled energy of IRAS at 100~$\mu$m. Our MIPS data indicate that extended emission is indeed present around UYSO\,1 on these scales. The projected distance between UYSO\,1 and HD~53623 is 0.46\,pc. Since the cloud is illuminated from the side, the real distance should not be much different, i.e., $<$1\,pc (if it protrudes less than 60\degr{} out of the plane of the sky). When seen from HD~53623, the size of the extended emission subtends an angle of 28--60$^\circ$, or a solid angle of 0.18--0.85~sr, assuming that the cloud is at a distance of 1--0.46\,pc and appears circular. The remaining uncertainty is the luminosity class of HD~53623 for which spectral types of B1V \citep{cla74} and B1II/III \citep{hou88} are given in the literature. Correspondingly assuming luminosities of $16000\,L_\odot$ or $39000\,L_\odot$ \citep{sch82}, the cloud core intercepts and eventually re-radiates 230--2640\,$L_\odot$. With the uncertainties involved, it seems reasonable to assume that the previously estimated upper limit for the luminosity was dominated by externally heated extended emission.

\begin{figure}
  \centering
\includegraphics*[width=6.0cm, angle=-90]{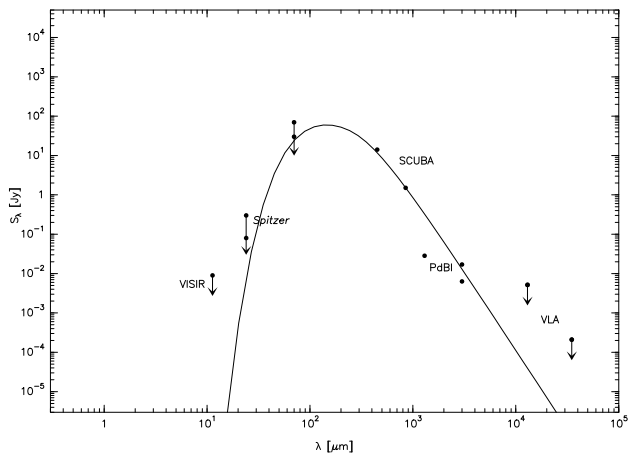}
  \caption{The SED of the combined two sources that make up UYSO\,1. The fit mainly relies on the submillimeter continuum data and the \textsl{Spitzer} upper limits and thus again constitutes only an upper limit in its infrared part. The two data points at 3~mm correspond to the high-resolution and low-resolution PdBI data. Fitting modified-blackbody radiation to this SED yields an upper limit for the estimated total luminosity of $\sim$50~$L_\odot$.}
  \label{uyso1SEDN.eps}
\end{figure}

In spite of the new data, the SED is only constrained by upper limits in the infrared regime. Since the newly discovered two components are not resolved in the submillimeter continuum, we use the unresolved PdBI continuum data at 3~mm (from the D configuration) and do not take into account the high-resolution PdBI data for the SED fit. Assuming an isotropic radiation field, a new modified-blackbody fit (with dust opacities of $\kappa_\nu \propto \nu^\beta$ and $\beta=2$) yields a luminosity upper limit of $\sim50\,L_\odot$ for both components combined (see Fig.~\ref{uyso1SEDN.eps}), assuming a distance of 1~kpc. This may be a severe lower bound if most of the luminosity escapes along the outflow cavities and is not re-processed into infrared radiation.

In spite of the uncertainties involved, it is interesting to compare this luminosity to an estimate of the accretion luminosity. The outflow mass entrainment rate translates into a mass loss rate of the driving jet which in turn translates into the actual accretion rate onto the star. \citet{bss02} estimate that this actual accretion rate is lower than the outflow mass entrainment rate by a factor of about 6. The above-mentioned conservative lower limit for the mass entrainment rate thus translates into a continuous accretion rate of $>5\times10^{-5}\,M_\odot\rm yr^{-1}$. A mass entrainment rate of $\dot{M}=4\times10^{-3}\,M_\odot$\,yr$^{-1}$ translates into an accretion rate of $7\times10^{-4}\,M_\odot$\,yr$^{-1}$. Based on the accretion rate, it is possible to estimate the accretion luminosity, $L_{\rm acc} = G \cdot M_\star \cdot \dot{M}_{\rm acc} \cdot r_\star^{-1}$. Of course, the stellar parameters are barely constrained in this case, and the situation is further complicated by the assumption of isotropic radiation, but we can carry out an order-of-magnitude check. For a 3~$M_\odot$ star and a corresponding radius of 5--8~$R_\odot$ \citep{pal92,yor08}, the lower limit of the accretion rate corresponds to an accretion luminosity of 600--900~$L_\odot$.
For comparison, a 10~$M_\odot$ star with a corresponding radius of 7--12~$R_\odot$ would have an accretion luminosity of 1300--2200~$L_\odot$. 
The estimated accretion luminosity thus appears to be about an order of magnitude larger than the luminosity upper limit deduced from the SED with the assumption of isotropic radiation. However, we note again that, given the likely presence of accretion disks, the radiation field is highly anisotropic. In particular, such disks would have the highest extinction, poorly determinable in the mid-infrared, in the direction towards the observer. Therefore, and due to the fact that the accretion luminosity is but a simple estimate with uncertain assumptions, the above discussion should be regarded as very tentative.

It remains unclear whether the empirical relation between the mass entrainment rate of the outflow and the luminosity of the central object that was used in \citetalias{jan04} is reliable (\citealp{shc96,hen00,bss02}, see also \citealp{wu05}). We note nevertheless that the mass entrainment rate appears to be higher than what would be expected when applying this relation for a driving source with a luminosity of $\sim$50~$L_\odot$.

UYSO\,1 was not observed to show centimetric NH$_3$ emission or millimetric N$_2$H$^+$ emission although we detect HCO$^+$(3-2) at the same position. \citet{wom91} find a similar effect towards Orion-IRc2 and speculate that N$_2$H$^+$ may be depleted by shocks or winds close to that source. On the other hand, \citet{taf04} found an extremely young starless core with constant C$^{18}$O and an unusually low N$_2$H$^+$ abundance. The fact that we do not see typical hot-core chemistry towards UYSO\,1, combined with the still low luminosity of the two sources suggests that we see very young sources. 
While it remains unclear how exactly the outflow fits into this picture due to the unknown inclination angle, a sufficiently large inclination angle would explain the extinction towards the central source and lead to a high mass entrainment rate of the outflow. 

\section{Conclusions}
\label{sec_conc}
We report the detection of two highly collimated jets intersecting close to the position of the previously identified candidate massive protostar UYSO\,1. Follow-up high-resolution millimeter radio observations show two continuum sources with masses of 3.5~$M_\odot$ (UYSO\,1a) and 1.2~$M_\odot$ (UYSO\,1b) and high column densities ($>10^{24}$~cm$^{-2}$) close to the geometric centers of the jets which probably are their powering sources. The projected distance between the two sources is 4200~AU. UYSO\,1a appears to power the energetic molecular outflow that was previously discovered towards UYSO\,1. UYSO\,1b contains a previously unknown H$_2$O maser source while a search for CH$_3$OH maser emission remained unsuccessful. The two millimeter protostars are not detected in deep near-infrared imaging and remain undetected also at mid- to far-infrared wavelengths. Also, no developing H\,{\sc II} regions were found. Submillimeter observations show that the region is rather cold and does not show typical hot-core chemistry. Curiously, not even emission in NH$_3$ or N$_2$H$^+$ was detected. An attempt to constrain the combined SED of the two sources yields an estimated luminosity of roughly $L\approx50\,L_\odot$, which is about an order of magnitude lower than a tentative estimate of the accretion luminosity. The key to combining these new results into a coherent picture probably lies in the unknown outflow inclination angle with respect to the line of sight. A large inclination angle would lead to configurations in which the circumstellar disk is seen nearly edge-on, obscuring the central object and helping to explain their infrared non-detections. For moderately large inclination angles of about $>25\,^\circ$, where a thick circumstellar disk and a dense envelope would obscure the central object, the molecular outflow powered by UYSO\,1a has a dynamical timescale of $<10000$ years with enormous outflow mass entrainment rates of $>\dot{M}=3\times10^{-4}\,M_\odot$\,yr$^{-1}$. The luminosity of the corresponding near-infrared jet is comparable to the most luminous jets found in a survey of Orion A. In addition to the peculiar chemistry and the still relatively low luminosity, this indicates a very early evolutionary stage.

\begin{acknowledgements} 
We wish to thank Sandra Bruenken for additional observations at Effelsberg, Jens Kauffmann for help with the 30m observations, Marian Szymczak for trying to detect CH$_3$OH emission towards UYSO\,1 at Toru\'n, Philippe Salom\'e and Robert Zylka, both at IRAM Grenoble,  for support in analyzing the PdBI data, as well as Henrik Beuther for helpful discussions. We would like to thank the referee, John Bally, for his helpful comments. T.S. is grateful for support through the A. v. Humboldt Foundation during his stay at the University of Hawaii. R.K. acknowledges support through Spitzer grant JPL\, no.\,1276999.
Partly based on Director's Discretionary Time observations collected at the Centro Astron\'omico Hispano Alem\'an (CAHA) at Calar Alto, operated jointly by the Max-Planck Institut f\"ur Astronomie and the Instituto de Astrof\'isica de Andaluc\'ia (CSIC). The National Radio Astronomy Observatory (NRAO) is operated by Associated Universities, Inc., under a cooperative agreement with the National Science Foundation. Partly based on observations carried out with the IRAM Plateau de Bure Interferometer and the IRAM 30m telescope. IRAM is supported by INSU/CNRS (France), MPG (Germany) and IGN (Spain). Partly based on observations made with ESO Telescopes at the Paranal Observatory under programme IDs 074.C-0648(A) and 076.C-0773(B). The James Clerk Maxwell Telescope is operated by The Joint Astronomy Centre on behalf of the Science and Technology Facilities Council of the United Kingdom, the Netherlands Organisation for Scientific Research, and the National Research Council of Canada. This work is based in part on observations made with the Spitzer Space Telescope, which is operated by the Jet Propulsion Laboratory, California Institute of Technology under a contract with NASA. Partly based on observations with the 100-m telescope of the MPIfR (Max-Planck-Institut f\"ur Radioastronomie) at Effelsberg.
\end{acknowledgements}


\bibliographystyle{aa} 
\bibliography{0598} 


\end{document}